\documentclass[dvips]{article}
\usepackage{xy} 
\xyoption{all} 
\CompileMatrices 

\newdimen\proofrulebreadth \proofrulebreadth=.05em
\newdimen\proofdotseparation \proofdotseparation=1.25ex
\newdimen\proofrulebaseline \proofrulebaseline=2ex
\newcount\proofdotnumber \proofdotnumber=3
\let\then\relax
\def\hfi{\hskip0pt plus.0001fil}
\mathchardef\squigto="3A3B
\newif\ifinsideprooftree\insideprooftreefalse
\newif\ifonleftofproofrule\onleftofproofrulefalse
\newif\ifproofdots\proofdotsfalse
\newif\ifdoubleproof\doubleprooffalse
\let\wereinproofbit\relax
\newdimen\shortenproofleft
\newdimen\shortenproofright
\newdimen\proofbelowshift
\newbox\proofabove
\newbox\proofbelow
\newbox\proofrulename
\def\shiftproofbelow{\let\next\relax\afterassignment\setshiftproofbelow\dimen0 }
\def\shiftproofbelowneg{\def\next{\multiply\dimen0 by-1 }%
\afterassignment\setshiftproofbelow\dimen0 }
\def\setshiftproofbelow{\next\proofbelowshift=\dimen0 }
\def\setproofrulebreadth{\proofrulebreadth}

\def\prooftree{
\ifnum  \lastpenalty=1
\then   \unpenalty
\else   \onleftofproofrulefalse
\fi
\ifonleftofproofrule
\else   \ifinsideprooftree
        \then   \hskip.5em plus1fil
        \fi
\fi
\bgroup
\setbox\proofbelow=\hbox{}\setbox\proofrulename=\hbox{}%
\let\justifies\proofover\let\leadsto\proofoverdots\let\Justifies\proofoverdbl
\let\using\proofusing\let\[\prooftree
\ifinsideprooftree\let\]\endprooftree\fi
\proofdotsfalse\doubleprooffalse
\let\thickness\setproofrulebreadth
\let\shiftright\shiftproofbelow \let\shift\shiftproofbelow
\let\shiftleft\shiftproofbelowneg
\let\ifwasinsideprooftree\ifinsideprooftree
\insideprooftreetrue
\setbox\proofabove=\hbox\bgroup$\displaystyle 
\let\wereinproofbit\prooftree
\shortenproofleft=0pt \shortenproofright=0pt \proofbelowshift=0pt
\onleftofproofruletrue\penalty1
}

\def\eproofbit{
\ifx    \wereinproofbit\prooftree
\then   \ifcase \lastpenalty
        \then   \shortenproofright=0pt  
        \or     \unpenalty\hfil         
        \or     \unpenalty\unskip       
        \else   \shortenproofright=0pt  
        \fi
\fi
\global\dimen0=\shortenproofleft
\global\dimen1=\shortenproofright
\global\dimen2=\proofrulebreadth
\global\dimen3=\proofbelowshift
\global\dimen4=\proofdotseparation
\global\count255=\proofdotnumber
$\egroup  
\shortenproofleft=\dimen0
\shortenproofright=\dimen1
\proofrulebreadth=\dimen2
\proofbelowshift=\dimen3
\proofdotseparation=\dimen4
\proofdotnumber=\count255
}

\def\proofover{
\eproofbit 
\setbox\proofbelow=\hbox\bgroup 
\let\wereinproofbit\proofover
$\displaystyle
}%
\def\proofoverdbl{
\eproofbit 
\doubleprooftrue
\setbox\proofbelow=\hbox\bgroup 
\let\wereinproofbit\proofoverdbl
$\displaystyle
}%
\def\proofoverdots{
\eproofbit 
\proofdotstrue
\setbox\proofbelow=\hbox\bgroup 
\let\wereinproofbit\proofoverdots
$\displaystyle
}%
\def\proofusing{
\eproofbit 
\setbox\proofrulename=\hbox\bgroup 
\let\wereinproofbit\proofusing
\kern0.3em$
}

\def\endprooftree{
\eproofbit 
  \dimen5 =0pt
\dimen0=\wd\proofabove \advance\dimen0-\shortenproofleft
\advance\dimen0-\shortenproofright
\dimen1=.5\dimen0 \advance\dimen1-.5\wd\proofbelow
\dimen4=\dimen1
\advance\dimen1\proofbelowshift \advance\dimen4-\proofbelowshift
\ifdim  \dimen1<0pt
\then   \advance\shortenproofleft\dimen1
        \advance\dimen0-\dimen1
        \dimen1=0pt
        \ifdim  \shortenproofleft<0pt
        \then   \setbox\proofabove=\hbox{%
                        \kern-\shortenproofleft\unhbox\proofabove}%
                \shortenproofleft=0pt
        \fi
\fi
\ifdim  \dimen4<0pt
\then   \advance\shortenproofright\dimen4
        \advance\dimen0-\dimen4
        \dimen4=0pt
\fi
\ifdim  \shortenproofright<\wd\proofrulename
\then   \shortenproofright=\wd\proofrulename
\fi
\dimen2=\shortenproofleft \advance\dimen2 by\dimen1
\dimen3=\shortenproofright\advance\dimen3 by\dimen4
\ifproofdots
\then
        \dimen6=\shortenproofleft \advance\dimen6 .5\dimen0
        \setbox1=\vbox to\proofdotseparation{\vss\hbox{$\cdot$}\vss}%
        \setbox0=\hbox{%
                \advance\dimen6-.5\wd1
                \kern\dimen6
                $\vcenter to\proofdotnumber\proofdotseparation
                        {\leaders\box1\vfill}$%
                \unhbox\proofrulename}%
\else   \dimen6=\fontdimen22\the\textfont2 
        \dimen7=\dimen6
        \advance\dimen6by.5\proofrulebreadth
        \advance\dimen7by-.5\proofrulebreadth
        \setbox0=\hbox{%
                \kern\shortenproofleft
                \ifdoubleproof
                \then   \hbox to\dimen0{%
                        $\mathsurround0pt\mathord=\mkern-6mu%
                        \cleaders\hbox{$\mkern-2mu=\mkern-2mu$}\hfill
                        \mkern-6mu\mathord=$}%
                \else   \vrule height\dimen6 depth-\dimen7 width\dimen0
                \fi
                \unhbox\proofrulename}%
        \ht0=\dimen6 \dp0=-\dimen7
\fi
\let\doll\relax
\ifwasinsideprooftree
\then   \let\VBOX\vbox
\else   \ifmmode\else$\let\doll=$\fi
        \let\VBOX\vcenter
\fi
\VBOX   {\baselineskip\proofrulebaseline \lineskip.2ex
        \expandafter\lineskiplimit\ifproofdots0ex\else-0.6ex\fi
        \hbox   spread\dimen5   {\hfi\unhbox\proofabove\hfi}%
        \hbox{\box0}%
        \hbox   {\kern\dimen2 \box\proofbelow}}\doll%
\global\dimen2=\dimen2
\global\dimen3=\dimen3
\egroup 
\ifonleftofproofrule
\then   \shortenproofleft=\dimen2
\fi
\shortenproofright=\dimen3
\onleftofproofrulefalse
\ifinsideprooftree
\then   \hskip.5em plus 1fil \penalty2
\fi
}

\usepackage{amssymb}
\usepackage{amstext}
\usepackage{amsmath}
\usepackage{amsthm}
\usepackage{rotating}

\newcommand{\id}{{\rm id}}
\newcommand{\Var}{{\it Var}}
\newcommand{\Prg}{\PPP}
\newcommand{\Runs}{\RRR}
\newcommand{\Prgg}{\PPp}
\newcommand{\Tr}{{Tr}}
\newcommand{\Intr}{\JJJ}
\newcommand{\Intt}{{\sf Int}}
\newcommand{\bv}{{\rm BV}}
\newcommand{\fv}{{\rm FV}}
\newcommand{\NSPK}{{\rm NSPK}}

\renewcommand{\to}{\xymatrix@C-.5pc{\ar[r]&}}
\newcommand{\ot}{\xymatrix@C-.5pc{& \ar[l]}}
\newcommand{\tto}[1]{\xymatrix@C-.5pc{\ar[r]^{#1}&}}
\newcommand{\oot}[1]{\xymatrix@C-.5pc{&\ar[l]_{#1}}}
\newcommand{\mono}{\xymatrix@C-.5pc{\ar@{>->}[r]&}} 
\newcommand{\epi}{\xymatrix@C-.5pc{\ar@{->>}[r]&}}
\newcommand{\mmono}[1]{\xymatrix@C-.5pc{\ar@{>->}[r]^{#1}&}} 
\newcommand{\eepi}[1]{\xymatrix@C-.5pc{\ar@{->>}[r]^{#1}&}}
\renewcommand{\mapsto}{\xymatrix@C-.5pc{\ar@{|->}[r]&}}
\newcommand{\mmapsto}[1]{\xymatrix@C-.5pc{\ar@{|->}[r]^{#1}&}}
\newcommand{\inclusion}{\xymatrix@C-.5pc{\ar@{^{(}->}[r] &}}
\newcommand{\iinclusion}[1]{\xymatrix@C-.5pc{\ar@{^{(}->}[r]^{#1}&}}

\newcommand{\crl}{\circlearrowleft}
\newcommand{\rar}{\rightarrow}

\newcommand{\send}[1]{\left<#1\right>}
\newcommand{\recv}[1]{\left(#1\right)}
\newcommand{\new}[1]{(\nu #1)}
\newcommand{\now}[1]{(\tau #1)}
\newcommand{\match}[2]{\left(#1 = #2\right)}
\newcommand{\isA}{\ {isA}\ }
\newenvironment{oup}[1]{\arraycolsep2pt \left<\begin{array}{#1}}{\end{array}\right>}
\newenvironment{inp}[1]{\arraycolsep2pt\left(\begin{array}{#1}}{\end{array}\right)}
\newcommand{\binp}[1]{\begin{inp}{#1}}
\newcommand{\einp}{\end{inp}}
\newcommand{\boup}[1]{\begin{oup}{#1}}
\newcommand{\eoup}{\end{oup}}

\newcommand{\AAA}{{\cal A}}

\newcommand{\GGG}{{\cal G}}

\newcommand{\JJJ}{{\cal J}}

\newcommand{\LLL}{{\cal L}}
\newcommand{\MMM}{{\cal M}}

\newcommand{\PPP}{{\cal P}}

\newcommand{\RRR}{{\cal R}}

\newcommand{\TTT}{{\cal T}}

\newcommand{\WWW}{{\cal W}}

\renewcommand{\Bbb}{\mathbb}

\newcommand{\CCc}{{\Bbb C}}

\newcommand{\IIi}{{\Bbb I}}

\newcommand{\LLl}{{\Bbb L}}

\newcommand{\NNn}{{\Bbb N}}

\newcommand{\PPp}{{\Bbb P}}

\mathcode`\<="4268 
\mathcode`\>="5269 
\mathchardef\gt="313E 
\mathchardef\lt="313C 

 \def\pushright#1{{
    \parfillskip=0pt            
    \widowpenalty=10000         
    \displaywidowpenalty=10000  
    \finalhyphendemerits=0      
    \leavevmode                 
    \unskip                     
    \nobreak                    
    \hfil                       
    \penalty50                  
    \hskip.2em                  
    \null                       
    \hfill                      
    {#1}                        
    \par}}                      

 \def\qed{\pushright{$\square$}\penalty-700 \smallskip}

\newenvironment{prf}[1]{\begin{trivlist} \item[{\bf ~Proof}#1.]}%
{\qed\end{trivlist}}

\newcommand{\be}[1]{\begin{#1}}

\newcommand{\ee}[1]{\end{#1}}
\newcommand{\beq}{\begin{equation}}
\newcommand{\eeq}{\end{equation}}
\newcommand{\ba}[1]{\begin{array}{#1}}
\newcommand{\ea}{\end{array}}
\newcommand{\bea}{\begin{eqnarray}}
\newcommand{\eea}{\end{eqnarray}}
\newcommand{\bear}{\begin{eqnarray*}}
\newcommand{\eear}{\end{eqnarray*}}
\newcommand{\bpr}{\begin{prf}{}}
\newcommand{\epr}{\end{prf}}
\newcommand{\bprf}[1]{\begin{prf}{#1}}
\newcommand{\eprf}{\end{prf}}

\theoremstyle{plain}
\newtheorem{proposition}{Proposition}[section]
\newtheorem{theorem}[proposition]{Theorem}
\newtheorem{lemma}[proposition]{Lemma}

\theoremstyle{definition}
\newtheorem{definition}[proposition]{Definition}

\theoremstyle{remark}
\newtheorem*{remarkno}{Remark}
\newtheorem*{questiono}{Question}

\usepackage{fullpage}

\title{Tracing the Man in the Middle\\ in Monoidal Categories}
\author{Dusko Pavlovic\\
Royal Holloway, University of London, and University of Twente\\
{\small Email:~dusko.pavlovic@rhul.ac.uk}
}

\date{}

\begin{document} 

\maketitle

\begin{abstract}
Man-in-the-Middle (MM) is not only a ubiquitous attack pattern in security, but also an important paradigm of network computation and economics. Recognizing  ongoing MM-attacks is an important security task; modeling MM-interactions is an interesting task for semantics of computation. Traced monoidal categories are a natural framework for MM-modelling, as the trace structure provides a tool to hide what happens \emph{in the middle}. An effective analysis of what has been traced out seems to require an additional property of traces, called \emph{normality}. We describe a modest model of network computation, based on partially ordered multisets (pomsets), where basic network interactions arise from the monoidal trace structure, and  a normal trace structure arises from an iterative, i.e. coalgebraic structure over terms and messages used in computation and communication. The correspondence is established using a convenient monadic description of normally traced monoidal categories.
\end{abstract}

\section{Introduction}
\subsubsection{Computation as interaction.}
If computers are viewed as state machines (e.g. Turing machines, or automata), then computations are their executions, i.e. sequences of actions, and one can reason about such computations in
terms of predicates over sequences of actions. Program correctness is established by proving that, for all possible executions, {\em bad things will not happen}, and that {\em good things will happen}. This is guaranteed, respectively, by the {\em safety\/} and the {\em liveness\/} properties \cite{Lamport77,Alpern-Schneider}.

Often, however, this simple view of computation needs to be refined to capture not only abstract actions, but also {\em locality\/} of data and controls, and the {\em interactions\/} that cause data
flows and control flows from one locality to another. This view of {\em computation as interaction\/} has been at the core of some later developments in program semantics \cite{Girard:GoI,AbramskyS:GoI,Abramsky:IC,PavlovicD:SIC}. One of its clearest and most prominent expressions has been game semantics of computation \cite{AbramskyS:GameSurvey,HylandOng}. With the Internet and computer networks, computation as interaction pervaded everyday life, and the network became the computer \cite{PavlovicD:CSR08}. Semantically, this means that computations cannot be reduced to linear sequences of abstract actions any more, i.e. that the latent information flows cannot be abstracted away. This is where security takes the center stage of computation: the new correctness requirement become that {\em bad information flows do not happen\/} and that {\em good information flows do happen}. The former roughly corresponds to the {\em secrecy\/} family of security properties (e.g. confidentiality, privacy, anonymity), whereas the latter corresponds to the {\em authenticity\/} family (integrity, non-malleability\ldots). But while the safety and the liveness properties where generally independent on each other, and in fact orthogonal (in the sense that each property can be uniquely decomposed into an intersection of a safety property and a liveness property \cite{Alpern-Schneider}), the secrecy and the authenticity properties usually depend on each other in complex and subtle ways, since every secret needs to be authenticated, and most authentications are based on secrets. Remarkably, one of the fundamental attack patterns on authentication protocols, which often goes under the name \emph{Man-in-the-Middle (MM)} \cite{Rivest-Shamir-eavesdropper,STS,Kelsey-Schneier-Wagner}, turns out to arise through deformations of the \emph{copycat strategy}, as the fundamental interaction pattern, modelling buffers, and supplying the identities in the interaction categories \cite{Abramsky:IC,AbramskyS:SoI}. In the present paper we formalize this observation. The ultimate goal is to provide a framework to trace back the buffer deformations, and thus trace the MM attacks.

\subsubsection{Tracing Man-in-the-Middle.}
We propose to apply categorical methods of semantics of interaction to security. The MM attack pattern, formalized in \emph{cord calculus}, originally designed for protocol analysis, naturally leads to a categorical \emph{trace structure}, generalizing the traces of linear operators in this case by means of a coalgebraic, iterative structure of the term algebra used in computation and communication. In the MM-attacks on authentication protocols, the intruder inserts himself\footnote{I hope that no one will be offended by the established genderism of the \emph{Man}\/-in-the-Middle terminology. For better or for worse, the "Man" is in concrete examples in the literature usually called Eve, or Carol.}  between the honest parties, and impersonates them to each other. MM is the strategy used by the chess amateur who plays against two grand masters in parallel, and either wins against one of them, or ties with both. MM is also used by the spammers, whose automated agents solve the automated Turing test by passing it to the human visitors of a free porn site, set up for that purpose \cite{DoctorowC:captcha-porn}. MM is, in a sense, one of the dominant business model on the web, where the portals, search engines and social networks on one hand insert themselves between the producers and the consumers of information, and retrieve freely gathered information for free, but on the other hand use their position in the middle to insert themselves between the producers and the consumers of goods, and supply advertising for a fee. In security protocols, MM is, of course, an interesting attack pattern. The fact that an MM attack on the famous Needham-Schroeder Public Key (NSPK) protocol \cite{Needham-Schroeder} remained unnoticed for 17 years promoted this toy protocol into what seemed to be one of the most popular subjects in formal approaches to security. Although the habit of mentioning NSPK in every paper formalizing security has subsided, we remain faithful to the tradition, and illustrate our MM-modeling formalism on the NSPK protocol. More seriously, though, the hope is that this formalism can be used to explore the scope and the power of the MM pattern in general, and in particular to formalize the idea of the \emph{chosen protocol attack}, put forward a while ago \cite{Kelsey-Schneier-Wagner}, but never explored mathematically.

\subsubsection{Background and related work.} The claim of the present paper is that the structure of the MM attacks can be faithfully presented and usefully analyzed in \emph{traced monoidal categories} \cite{JSV}. The syntactic trace structure of the constructed categories is used to \emph{trace out} the intruder, just like, e.g., the linear trace structure of complex vector spaces is used to trace out the \emph{ancillae} in quantum systems. The central technical feature is that the trace structure of the particular MM frameworks arises from the \emph{iterative} structure \cite{ElgotC:iterative,Bloom-Esik:book,MossL:parametric,AczelP:iterative} of the message algebras, which in effect resolves the term equations induced by the interactions, and thus propagates the data sent in messages. The coalgebraic nature of such iterative structures has been explained and analyzed in \cite{AdamekJ:free,AdamekJ:survey-coalg}, where also the further references can be found. The proposed framework for the MM interactions is built as an action category \cite{Milner:action,Milner:calculi} along the lines of \cite{PavlovicD:MSCS97}  from the \emph{cord calculus} for protocol analysis \cite{PavlovicD:CSFW01,PavlovicD:JCS04,PavlovicD:JCS05,PavlovicD:ESORICS06}, which was designed as a domain specific process calculus underlying an integrated development environment for security protocols \cite{PavlovicD:ARSPA06}. More detailed explanations will be provided in the text, as the formalism is introduced.

\subsubsection{Outline of the paper.} Cord calculus is described in Sec.~\ref{Processes-sec}, and arranged into a suitable categorical structure. Categorical semantics of the MM-interactions is described and analyzed in Sec.~\ref{Interactions-sec}. The example of the MM-attack on the NSPK protocol is worked out in Sec.~\ref{Protocols-sec}. Finally, Sec.~\ref{Discussion-sec} discusses the presented approach and some ideas for future work. The categorical background (some of it apparently of independent interest) is presented in three appendices. 

\section{Cord semantics of processes}\label{Processes-sec}
In this section we introduce cord spaces and build cord categories. Various
versions of the cord formalism were used in
\cite{PavlovicD:CSFW01,PavlovicD:MFPS03,PavlovicD:FMSE03,PavlovicD:CSFW03,PavlovicD:CSFW04,PavlovicD:JCS04,PavlovicD:JCS05}. It was a simple reaction-based process calculus, obtained by extending the strand space formalism
\cite{strands} by variables and a substitution mechanism, capturing
the information flows (e.g., in protocols where participants forward parts of a payload encrypted by someone else's public key). The current version simplifies away the particle reactions, and separates the term substitution mechanism from the partial ordering of actions. The latter part remains  close in spirit to strand spaces, or to Lamport's preorders \cite{Lamport:Time}, which can be viewed as a predecessor of all such formalisms. Formally, all such formalisms subsume under Pratt's \emph{p}\/artially \emph{o}\/rdered \emph{m}\/ulti\emph{sets} (pomsets) \cite{Pratt:pomsets,Gischer}. In the versions from \cite{PavlovicD:ESORICS04,PavlovicD:CSFW05,PavlovicD:ESORICS06,PavlovicD:dist06}, a cord space is thus simply a pomset of actions with {\em localities}, i.e. distributed among distinct agents. To represent communication, the actions include sending and receiving messages. The messages are terms of a polynomial algebra, supporting variable assignment and substitution. The most recent version is in \cite{PavlovicD:ICDCIT12}.

\subsection{Cord spaces and their runs}\label{Cord spaces}
Processes are built starting from abstract sets of
\begin{itemize}
\item\textbf{terms} $\TTT$, with enough variables $\Var_\TTT\subseteq \TTT$,
\item\textbf{agents (or locations)} $\WWW$, with $\Var_\WWW\subseteq \WWW$ and
\item\textbf{actions} $\AAA$, which comes with the constructors such as 
\[
\WWW^2 \times \TTT \tto{\send{-}}  \AAA\qquad\qquad
\Var_\WWW^2 \times \Var_\TTT \tto{\recv{-}}  \AAA \qquad \ldots
\]
that generate at least the send actions $\send{A\rar B:t}$ and the
receive actions $\recv{X\rar Y:z}$, and moreover other actions which a
particular model may require, such as $\new{x}$, $\now{x}$,
$\match{x}{t}$, or $\left(t/p(x)\right)$
\cite{PavlovicD:CSFW05,PavlovicD:ARSPA06}.
\end{itemize}
A {\bf cord space} $P$ is a map $\LLl \tto{\isA} \AAA\times \WWW$,
where the set of {\em actions\/} $\LLl = \LLl_P$ comes equipped with a
preorder $\leq$, representing their {\em temporal\/} ordering. Recall that a preorder is a transitive and reflexive
relation. The set $\LLL$ of cord spaces carries two monoid structures:
\begin{itemize}
\item $(\LLL,\otimes,\emptyset)$, where $P\otimes Q: \LLl_P + \LLl_Q
\xymatrix@C-.3pc{\ar[rr]^{[isA_P, isA_Q]}&&} \AAA\times \WWW$ is the
cord space over the preorder $\LLl_P + \LLl_Q$ where the actions of $P$ remain incomparable with the
actions of $Q$, and 
\item $(\LLL,\cdot,\emptyset)$, where $P \cdot Q: \LLl_P \lt \LLl_Q
\xymatrix@C-.3pc{\ar[rr]^{[isA_P, isA_Q]}&&} \AAA\times \WWW$ is the cord space over the preorder $\LLl_P \lt \LLl_Q$ where every action of $P$ precedes every action of $Q$. 
\end{itemize}
In each case, $\emptyset$ represents the empty cord space $\emptyset \to \AAA\times \WWW$. Clearly, these two
operations respectively correspond to the parallel and the sequential
composition of cord spaces. Repeated application of these operations
to actions generates most, but not all cord spaces \cite{Gischer}. Given a cord space $P$, its sets of the receive and the send actions are
\bear
{\sf recvs}(P) & = & \{\ell \in \LLl_P\ |\ \exists XYz.\ \ell \isA \recv{X\rar Y:z}\}\\
{\sf sends}(P) & = & \{\ell \in \LLl_P\ |\ \exists ABt.\ \ell \isA \send{A\rar B:t}\}
\eear
A {\bf run}  of the cord space $P$ is a map $\surd_P : {\sf recvs}(P) \to {\sf sends}(P)$ such that
\bear
k = \surd \ell & \Longrightarrow & k \not\geq \ell
\eear
In other words, extending the temporal preorder by setting for every
$\ell \in {\sf recvs}(P)$ that $\surd \ell \leq \ell$ must not
introduce any new cycles. 

\begin{remarkno} The temporal ordering of $\LLl$ is not required to
be asymmetric, because different actions $p\neq q$ may occur at the
same time, and thus satisfy $p\leq q$ and $p \geq q$. With abstract
actions, one could assume that such actions can be identified, or
sequentialized. This is done in Pratt's pomsets (partially ordered
multisets) \cite{Pratt:pomsets}. However, when actions involve terms,
as they do in the above model, and when an action may depend on
another action for the values that need to be substituted before it
can be executed, then the temporal precedence loops may correspond to
{\em deadlocks}. Effective runs, of course, need to be
deadlock-free. Even if the notion of a cord space was restricted to
disallow temporal loops, such loops would arise as deadlocked runs,
and would need to be taken into account. E.g., a cord space with a
single send and a single receive action has no runs if the sent terms
depend on some received data. Cord Spaces with no runs, and with
cyclic dependencies arise naturally, and the existence of effective
runs cannot be imposed, but needs to be analyzed. 
\end{remarkno}

\subsection{Cord processes}\label{Cord processes}
\subsubsection{From action structures to interaction categories.}
The composition operations over cord spaces naturally lead to a categorical structure, as soon as the input and the output interfaces of cords are displayed. The categorical composition of the cord processes can be obtained from the sequential composition of cord spaces, whereas the parallel composition yields the monoidal structure. The resulting category can be viewed as an instance of Milner's {\em action structure\/} construction \cite{Milner:action,Milner:calculi}. This view uncovers a common structural denominator for a wide gamut of process representations. Moreover, it provides a uniform framework for the categorical abstraction operations \cite{PavlovicD:MSCS97}, which we shall use to capture secure information flows.\footnote{Our low level syntax, with the convention that input interface and the binding operators are denoted by the round brackets $\binp{c} \vec{x}\einp$, whereas the output interface and the send action are written in the angle brackets $\boup{c}\vec{s}\eoup$, is inherited from the action calculus. For the high level structures, though, the  graphic notation for monoidal, traced and compact categories is more convenient, and will be mixed with the syntactic calculus.}

However, the cord category presented here is, strictly speaking, {\em not\/} an action structure. Although its objects, its morphisms and even its syntactic presentation are just as in an action structure, its composition is not derived from the sequential composition of processes, but rather from a minimal sequentialisation of the parallel composition. The upshot of this is that the resulting category  carries a natural {\em trace\/} structure \cite{JSV}, in contrast with the original action structures. This structure will then be used to define the composition in a {\em category of interactions}. Security protocols can be specified as certain interactions, i.e. morphisms in that category; certain attacks on them then arise by composing interactions.

The idea that the composition of interactions can be characterized as a combination of {\em parallel composition and hiding} has previously been developed within the framework of {\em interaction categories\/} \cite{Abramsky:IC}. The fact that the present setting requires deviating from the sequential composition, and defining a categorical composition based on the parallel composition and hiding, in order to support the trace structure, needed for modeling interactions --- can be viewed as a confirmation, and an interesting realisation of that idea.

\subsubsection{Simple typing.}\label{Simple typing}
A {\bf cord process} consists of a cord space with input and output
declarations. If we assume, for simplicity, that all data are of the
same type, then an input interface declaration boils down to a tuple
of distinct variables $\vec{x}_{1..\ell} =\binp{cccc} x_1 & x_2 &
\ldots & x_\ell\einp$, and the output interface is a tuple arbitrary
terms $\vec{s}_{1..m}=\boup{cccc} s_1 & s_2 & \ldots &
s_m\eoup$. Either of these tuples can be empty, in which case we write
() and $<>$. A cord process $p:\ell \to m$ is thus viewed as an expression
in the form $\binp{c}\vec{x}_{1..\ell}\einp [P]
\boup{c}\vec{s}_{1..m}\eoup$, where $P$ is a cord space. The variables
$\vec{x}_{1..\ell}$ may or may not occur in (the terms of the actions
of) $P$ and in $\vec{s}_{1..m}$. More precisely, a cord process $p:\ell\to
m$ is an equivalence class of such expressions modulo variable
renaming, consistently throughout $P$ and $\vec{s}_{1..m}$.

The category $\Prg$ consists of arities and the cord processes between them. It will often be convenient to extend the above action calculus syntax
to diagrams, with the cord spaces enclosed in the boxes, and the
interfaces displayed on the arrows: 
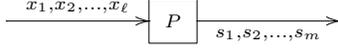
\[\def\objectstyle{\scriptstyle} 
\xymatrix{
\ar[rr]^-{x_1,  x_2, \ldots, x_\ell} &&
*+<1pc>[F]{P} \ar[rr]_-{s_1, s_2, \ldots,  s_m} &&
}
\]
The {\bf composition} of cord processes
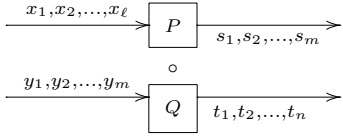
\[\def\objectstyle{\scriptstyle} 
\xymatrix@R-1.8pc{
\ar[rr]^-{x_1,  x_2, \ldots, x_\ell} &&
*+<1pc>[F]{P} \ar[rr]_-{s_1, s_2, \ldots,  s_m} &&\\ 
&& \circ \\ 
\ar@<1ex>[rr]^-{y_1,  y_2, \ldots, y_m} &&
*+<1pc>[F]{Q} \ar@<1ex>[rr]_{t_1,t_2,\ldots,t_n} &&
}
\]
can then be obtained by connecting the output interface of $P$ with the input interface of $Q$
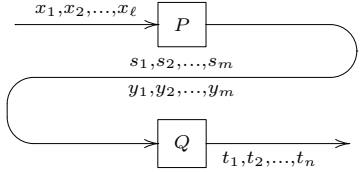
\[\def\objectstyle{\scriptstyle} 
\xymatrix@R-1.8pc{
\ar[rr]^-{x_1,  x_2, \ldots, x_\ell} &&
*+<1pc>[F]{P} \ar `r[rr] `[d] 
`[ddddll]_-{s_1, s_2, \ldots,  s_m}^-{y_1,  y_2, \ldots, y_m} `[dddd] [dddd] &&\\ 
&&  \\
&& \\
&& \\
&&
*+<1pc>[F]{Q} \ar[rr]_{t_1,t_2,\ldots,t_n} &&
}
\]
and performing the induced substitutions:
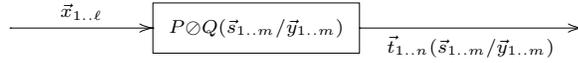
\[
\def\objectstyle{\scriptstyle} \xymatrix{
\ar[rr]^-{ \vec{x}_{1..\ell}} && *+<1pc>[F]{P\oslash Q(\vec{s}_{1..m}/\vec{y}_{1..m})} \ar[rrr]_-{ \vec{t}_{1..n}(\vec{s}_{1..m}/\vec{y}_{1..m})} &&&
}
\]
where the cord space $P\oslash Q(\vec{s}_{1..m}/\vec{y}_{1..m})$ is the minimal order extension the parallel composition $P\otimes Q$, induced by the substitution $(\vec{s}_{1..m}/\vec{y}_{1..m})$. To understand why the order needs to be extended, note that substituting, say $s_i(z)$ for $y_i$ in an action $b$ of $Q$ may introduce a variable $z$, which may be bound to an action $a \isA \recv{z}$ in $P$. In such cases, we must add $a\lt b$ to the preorder of $P\oslash  Q(\vec{s}_{1..m}/\vec{y}_{1..m})$ in order to preserve the information flow through $z$. 

The preorder $\LLl_P \oslash \LLl_Q$ of the cord space $P\oslash Q : \LLl_P \oslash \LLl_Q \to \AAA\times \WWW$ is thus defined over the underlying set $\LLl_P + \LLl_Q$, by setting
\begin{eqnarray}
a\lt b & \iff & a,b\in \LLl_P \wedge a \lt b \notag\\
& \vee & a,b\in \LLl_Q \wedge a \lt b \notag\\
& \vee & a\in \LLl_P \wedge b\in \LLl_Q \wedge \bv (a) \cap \fv (b) \neq \emptyset \label{lt}
\end{eqnarray}
where $\bv (a)$ denotes the set of variables bound by the action $\alpha$, such that  $a \isA \alpha $, and $\fv(b)$ is the set of the free variables that occur in the terms of the action $\beta$, such that $b\isA \beta$.\footnote{For a term $t\in \TTT$, the set $\fv(t)$ is defined by the usual inductive clauses. For an action $\alpha\in \AAA$, the set $\bv(\alpha)$ is just $\bv\recv{x} = \bv\new{x} =\{x\}$, and $\bv\match{x_1,\ldots,x_k}{t_1,\ldots,t_k} =\{x_1,\ldots,x_k\}$.}  

\begin{remarkno} The output terms are generally in the form $s_i = s_i(\vec{x},\vec{u},\vec{v})$, i.e. they may depend on any variables from three disjoint tuples:
\begin{itemize}
\item $\vec{x}$, which are bound to the input interface $\binp{c} \vec{x}\einp$,
\item $\vec{u}$,  which are bound to some binding operations on the cord space $P$, such as $\recv{u_i}$, $\new{u_i}$, or $\match{u_i}{t}$, and
\item $\vec{v}$ are free.
\end{itemize}
Any name clashes of any of these variables with any of the corresponding tuples of variables used in $Q$ and $t_j$ must be eliminated by renaming prior to the composition.
\end{remarkno}

The {\bf identities} $\id: n\to n$ in $\Prg$ can be thought of as the {\em
buffers}, i.e. the trivial cord processes that perform no processing, and just pass the inputs to the output interface. Formally, they are the expressions in
the form $\binp{ccc} x_1 & \ldots & x_n\einp[] \boup{ccc} x_1 & \ldots
& x_n\eoup$, where $[]$ is the empty cord space. In fact, every
function $f : \{1,2,\ldots,n\}\to \{1,2,\ldots m\}$ induces (contravariantly!) a unique
cord process $\pi_f : m\to n$, defined
\bear \pi_f & = & \binp{cccc} x_1 & x_2 &  \ldots & x_m\einp[]
\boup{cccc} x_{f(1)} & x_{f(2)} & \ldots & x_{f(n)}\eoup 
\eear
which just rearranges the data from the input, and displays them at
the output, with no processing. If $\NNn$ is the category of finite
sets $n = \{0,1,\ldots,n-1\}$ and functions between them, then $\pi :
\NNn\to \Prg$ can be viewed as a faithful functor, identity on
objects, thus displaying $\NNn$ as a subcategory of $\Prg$. As
explained in \cite{PavlovicD:MSCS97}, $\Prg$ is freely generated over
$\NNn$ by adjoining variables, terms, and cord spaces. 

If $\NNn$ itself is viewed as the free strictly cocartesian category
(i.e. with strict coproducts) over one generator, then $\Prg$ is the
free strictly monoidal category, generated by a single object, and the
morphisms induced by the terms from $\TTT$, and the cord spaces from
$\LLL = \LLL_{\TTT,\WWW,\AAA}$. The {\bf tensor product}
in $\Prg$ is induced by the parallel composition of cord spaces, and
the juxtaposition of the interfaces:  
\[\def\objectstyle{\scriptstyle} 
\xymatrix@R-2pc{
& \ar[rr]^-{x_1,  x_2, \ldots, x_\ell} &&
*+<1pc>[F]{P} \ar[rr]_-{s_1, s_2, \ldots,  s_m} &&\\ 
&&& \otimes \\ 
& \ar[rr]^-{x'_1,  x'_2, \ldots, x'_{\ell'}} &&
*+<1pc>[F]{P'} \ar[rr]_-{s'_1, s'_2, \ldots,  s'_{m'}} &&\\ \\
&&& = \\ \\
\ar[rrr]^-{x_1, \ldots, x_\ell, x'_1,\ldots,x'_{\ell'}} &&&
*+<1pc>[F]{P\otimes P'} \ar[rrr]_-{s_1, \ldots,  s_m, s'_1,\ldots, s'_{m'}} &&&
}
\]
where the variables from $\vec{x}_{1..\ell} = \binp{ccc} x_1&\ldots &
x_\ell \einp$ are (if necessary, renamed to be) disjoint from the
variables in $\vec{x'}_{1..\ell'} = \binp{ccc}x'_1&\ldots & x'_{\ell'}
\einp$, and the cord space  $P\otimes P'$ is the
parallel composition of cord spaces, where each action of $P$ remains
incomparable with all actions of $P'$. Clearly, the tensor $\otimes$
in $\Prg$ extends the coproduct $+$ in $\NNn$, and boils down to it on
the objects, i.e. $m\otimes n = m+n$. The tensor unit is thus the
empty arity 0. The obtained monoidal structure is symmetric, and
strictly associative and unitary.

\begin{remarkno} An action structure over cord spaces would be a monoidal category with the same objects and morphisms as above, and even the same monoidal structure. However, the composite of $\binp{c}\vec{x} \einp [P] \boup{c}\vec{s}\eoup$ and $\binp{c}\vec{y} \einp [Q] \boup{c}\vec{t}\eoup$ would be $\binp{c}\vec{x} \einp [P\cdot Q(\vec{s}/\vec{y})] \boup{c}\vec{s}\eoup$, rather than $\binp{c}\vec{x} \einp [P\oslash Q(\vec{s}/\vec{y})] \boup{c}\vec{s}\eoup$. 
\end{remarkno}

\begin{questiono} What is the universal property of  $\Prg = \Prg_\LLL$?  The general results of \cite{PavlovicD:MSCS97} tell that the action structure constructed over cord spaces is the free symmetric strictly monoidal category generated over a single object $1$, by adjoining
\begin{itemize} 
\item for every cord space $P\in \LLL$ an endomorphism $()[P]<>:0\to 0$, and 
\item for every variable $x$, an indeterminate arrow $()[]<x>:0\to 1$, and
\item an abstraction $(x)[]<>:1\to 0$.
\end{itemize} 
How does the modified composition of the cord category  $\Prg = \Prg_\LLL$ change this result? 
\end{questiono}

\subsubsection{Refined typing.}
For simplicity, in the above description of the category $\Prg$ we
ignored the issues of typing: an arity was just  a tuple. In order to
describe {\em distributed\/} cord processes, one must distinguish at the
interfaces the entries for the terms from the entries for agent identifiers. At the very least, the input and the output interfaces
thus need to be typed as products of the types $\TTT$ and $\WWW$ of
terms and agents respectively. Assuming that these two types do not
depend on each other in any way, the arities just split in two parts,
and become pairs of natural nubmers. A cord process $p: <k,m>\to <\ell,n>$
is now in the form 
\[ \binp{cc} \vec{X}_{1..k} & \vec{x}_{1..m} \einp [P] \boup{cc}
\vec{A}_{1..\ell} & \vec{s}_{1..n} \eoup 
\]
where $X_i$ are agent variables (roles), $x_i$ are term variables,
$A_i$ are agent identifiers (constant or variable), and $s_i$ are
arbitrary terms. 

Various aspects of distributed computation induce further refinements
of the type system. E.g. local variables, available only to particular
agents, are given with a map $\Var_\TTT \to \WWW$. A local variable
can thus be assigned a value only after its locality is known. This
means that in the input interface, it can only occur after the
corresponding agent variable. The typing of terms is thus dependent on
the type of agents, and we have at least one level of dependent type
theory. Further type dependencies arise because 
\begin{itemize}
\item an agent $A^\WWW[X^\WWW]$ may depend on the different roles $X$
that she may play in a cord space, 
\item a term $t^\TTT[X^\WWW]$ may be computed in different ways in
different roles, 
\item a term $t^\TTT[x^\TTT]$ may be computed in different ways
depending on the outcome of the computation of some other term\ldots 
\end{itemize}
In the present paper, we shall mostly need just the dependency of
terms on agents.  
In the rest of the paper, we denote by $\Gamma,\Phi, \Psi$ the general
types, as the objects of $\Prg$, in contrast with the simple arities
$\ell, m, n$. 

\subsubsection{Contexts.}\label{Contexts}
Let $\vec{v}:\Gamma$ be the tuple of all free variables that occur in
a cord process $p:\Phi\to \Psi$. Then $\Gamma$ is the {\em type context\/}
of the cord process $p$. The subcategory of $\Prg$ consisting of cord processes like $p$, whose free variables are contained in $\vec{v}$, is
denoted $\Prgg[\vec{v}:\Gamma]$, or just $\Prgg[\vec{v}]$. The
cord processes that contain no free variables constitute the category $\Prgg
= \Prgg[]$. It is easy to see that each $\Prgg[\vec{v}]$ is closed
under the monoidal structure of $\Prg$.  Alternatively, $\Prgg[\vec{v}]$ can be viewed as the subcategory of
$\Prgg$ spanned by the cord processes in the form $\binp{cc} \vec{v} & \vec{x}\einp\ [P]\ \boup{cc} \vec{v} &\vec{s} \eoup$.

\subsection{Runs of cord processes}\label{Runs}
A run of a cord process represented by the expression $\binp{c}\vec{x}\einp [P] \boup{c}\vec{s}\eoup$ is an expression in the form $\binp{c}\vec{x}\einp [P^\surd] \boup{c}\vec{s}\eoup$, where $P^\surd$ is a run extending the cord space $P$, i.e. a pair 
\bear
P^\surd & = &  \left(P, \surd_P : {\sf recvs}(P) \to {\sf sends}(P)\right)
\eear
The category  $\Runs$ of cord runs inherits all structure from the category $\Prg$ of cord processes, and comes with the obvious identity-on-the-objects forgetful functor $\Runs \to \Prg$.  

\begin{lemma}
Let $\surd_P : {\sf recvs}(P) \to {\sf sends}(P)$ and $ \surd_Q : {\sf recvs}(Q) \to {\sf sends}(Q)$ be runs of cord processes $\binp{c}\vec{x}\einp [P] \boup{c}\vec{s}_{1..k}\eoup$ and $\binp{c}\vec{y}_{1..k}\einp [Q] \boup{c}\vec{t}\eoup$. Then the disjoint union $\surd_P + \surd_Q : {\sf recvs}(P+Q) \to {\sf sends}(P+Q)$ is a correct run of the composite process $\binp{c}\vec{x}\einp [P\oslash Q(\vec{s}/\vec{y})] \boup{c}\vec{t}(\vec{s}/\vec{y})\eoup$.
\end{lemma}

\bpr
The correctness requirement of a run is that $\surd a \not\geq a$. This property will be preserved, because $\surd a$ is in the same component ($P$ or $Q$) as $a$, and the composition $P\oslash Q(\vec{s}/\vec{y})$ only makes some $Q$-actions come after some $P$ actions (that bind their variables). 
\epr

\section{Cord semantics of interactions}\label{Interactions-sec}
The informal idea of {\em process interaction\/} is that two processes feed each other some data, and process them together, i.e. partially evaluate over them. Two cord processes thus interact when a part of the outputs of one of them is piped to the input interface of the other, and vice versa.  This framework allows us to formalize some security questions: {\em What properties of the information flows are preserved under the interaction? Which new information flows emerge, and which old ones vanish?}  To formalize this idea, we partition the interfaces of each process into an Initiator and a Responder part. Two interaction programs are then composed by passing the Responder data from one to the Initiator interface of the other. These data are then propagated, and the abstract partial evaluation over them is performed using the trace structure of the cord category. 

\subsection{Iterations and traces of cord processes and runs}
Iterative algebras and their various extensions provide a widely
studied view of program execution
\cite{ElgotC:iterative,Bloom-Esik:book,MossL:parametric,AczelP:iterative}.
On the other hand, an important categorical form of iteration is given
by the notion of {\em traced\/} monoidal categories \cite{JSV}. In
categories of programs, these two structures turn out to coincide. 

\be{proposition}\label{proposition}
The uniform normal trace structures on the cord category $\Prg=\Prg_{\TTT,\WWW,\AAA}$ are in one-to-one correspondence with the iterative structures of the term algebra $\TTT$. 
\ee{proposition}

\subsubsection{Background.} The iterative structures are defined in \ref{iterative}. An elegant coalgebraic account of the various versions and extensions of iterative algebras can be found in \cite{AdamekJ:free}. A general survey of coalgebra from this angle is provided in \cite{AdamekJ:survey-coalg}. The general trace structures over monoidal categories are defined in \cite{JSV}. The cord category naturally carries a \emph{normal} trace structure. The normality requirement $\Tr^U(f\otimes U) = f$ is shared by the traces over relations, but not by the traces over vector spaces. The importance of this requirement in the current context is that it allows a functorial presentation of the trace operations, which seems crucial for detecting the MM attacks. A convenient algebraic characterization of normal traces is given in the Appendix.    

\bpr
Suppose that $\TTT$ is an iterative algebra. By Def.~\ref{iterative}, this means it contains the iteration operation $(-)^\dag$, which assigns to every system of $ k\leq \ell$ {\em guarded\/} equations 
\bear
\vec{y}_{1..k} & = & \vec{f}_{1..k}(\vec{y}_{1..\ell})\ \mbox{ has a unique solution }\\
\vec{f}^{\ \dag}_{1..k}(\vec{y}_{k+1..\ell}) & = & \boup{ccc} f^\dag_1(y_{k+1}, \ldots, y_\ell) &  \ldots & f^\dag_k(y_{k+1},\ldots,y_\ell) \eoup \mbox{ i.e. }\\
\vec{f^\dag}_{1..k}(\vec{y}_{k+1..\ell}) & = & \vec{f}_{1..k}\left(\vec{f^\dag}_{1..\ell}(\vec{y}_{k+1..\ell})\right)
\eear
The assumption that the equations are guarded means that the operations $\vec{f}_{1..k} = \boup{cccc} f_1 & f_2 & \ldots & f_k \eoup$ are not projections in the form $f_j(\vec{y}_{1..\ell}) = y_i$ for $1\leq i\leq k$. The operations different from such projections are called {\em guards}.

Using the iteration $(-)^\dag$, we define the trace functor $\Tr : \Prg^\crl \to \Prg$, i.e. the trace operation $\Tr^\ell_{mn}: \Prg(m\otimes \ell, n\otimes \ell) \to \Prg(m,n)$. The idea is that the trace $\Tr(p):m\to n$ of a program $p:m\otimes \ell\to n\otimes \ell$, which is in the form
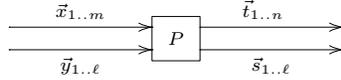
\[
\def\objectstyle{\scriptstyle} \xymatrix{
  \ar@<1ex>[rr]^-{ \vec{x}_{1..m}} \ar@<-1ex>[rr]_-{ \vec{y}_{1..\ell}} &&
*+<1pc>[F]{P} \ar@<1ex>[rr]^{\vec{t}_{1..n}} \ar@<-1ex>[rr]_-{\vec{s}_{1..\ell}} && 
}
\]
should be obtained by passing the outputs $\vec{s}_{1..\ell}$ to the inputs $\vec{y}_{1..\ell}$ 
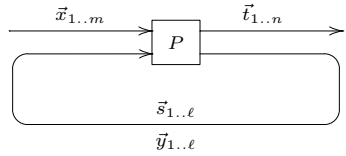
\[
\def\objectstyle{\scriptstyle} \xymatrix{
  \ar@<1ex>[rr]^-{ \vec{x}_{1..m}} &&
*+<1pc>[F]{P} \ar@<1ex>[rr]^{\vec{t}_{1..n}} \ar@<-1ex> `r[rr] 
`[d] 
`[ll]_-{\vec{s}_{1..\ell}}^-{ \vec{y}_{1..\ell}} `[] []   && 
\\
&&&&}
\]
However, since the variables $\vec{y}_{1..\ell}$ may occur in the  terms $\vec{s}_{1..\ell}$, this substitution must be done iteratively: the terms  $\vec{s}_{1..\ell}$ must also be substituted for $\vec{y}_{1..\ell}$ in themselves. This iteration can terminate if the system of equations 
\bear
\vec{y}_{1..\ell} & = & \vec{s}_{1..\ell}(\vec{x},\vec{y}_{1..\ell},\vec{u},\vec{v})
\eear 
has a solution. Like before, we denote by $\vec{u}$ the variables that are bound to some binding operations in $P$, and by $\vec{v}$ the  free variables that occur in some $s_i$. In general, this system may not be guarded, i.e. some of the equations may boil down to the form $y_i = y_j$. With no loss of generality, we can rearrange the system so that the first $k$ equations $\vec{y}_{1..k} = \vec{s}_{1..k}(\vec{y}_{1..k},\vec{y}_{k..\ell},\vec{x}_{1..m})$ (for $0\leq k \leq \ell$) are guarded, whereas the last $\ell - k$ equations are in the form $y_i = y_j$.  Since the last $\ell - k$ equations just partition the variables $\vec{y}_{k..\ell}$, these equations can be eliminated by choosing a single representative for each equivalence class of variables. This yields a vector of variables $\vec{y}^{\ *}_{k..\ell}$, some of which may be equal.  On the other hand, the assumption that $\TTT$ is an iterative algebra implies that the guarded system $\vec{y}_{1..k} = \vec{s}_{1..k}(\vec{x},\vec{y}_{1..k},\vec{y}^{\ *}_{k..\ell},\vec{u},\vec{v})$ has a unique solution $\vec{s}^{\ \dag}_{1..k}(\vec{x},\vec{y}^{\ *}_{k..\ell},\vec{u},\vec{v})$, which means that $ \vec{s}^{\ \dag}_{1..k} = \vec{s}_{1..k}(\vec{x},\vec{s}^{\ \dag}_{1..k},\vec{y}^{\ *}_{k..\ell},\vec{u},\vec{v})$ holds.
We can now define the trace program $\Tr^\ell_{mn}(p):m\to n$ by substituting the solutions $\vec{s}^{\ \dag}_{1..k}$ and $\vec{y}^{\ *}_{k..\ell}$ for $\vec{y}_{1..\ell}$
\[
\def\objectstyle{\scriptstyle} \xymatrix{
  \ar[rr]^-{ \vec{x}_{1..m}}  &&
*+<1pc>[F]{P^\dag(\vec{s}^{ \dag}_{1..k},\vec{y}^{\ *}_{k..\ell}/\vec{y}_{1..\ell})} \ar[rrrr]^-{\vec{t}_{1..n}(\vec{s}^{ \dag}_{1..k},\vec{y}^{\ *}_{k..\ell}/\vec{y}_{1..\ell})} &&& &
}
\]
After these substitutions, it may occur that a variable $u_i$, bound in an action $\alpha$ of $P$ (e.g., a receive, or nonce generation action), may be introduced in the substitution instance of $\beta(\vec{s}^{\ \dag}/\vec{y})$ of some other action $\beta$ of $P$. Since the variable bindings implement the information flow, $\alpha$ must precede $\beta$ in $P^\dag$.

The cord space $P^\dag : \LLl^\dag \to \AAA\times \WWW$ is thus obtained from $P : \LLl \to \AAA\times \WWW$ by strengthening the ordering of $\LLl$ to capture this. We define $\LLl^\dag$ to be the preorder with the same elements as $\LLl$, and such that for all $a,b$ holds
\bear
a\lt b  \mbox{ in } \LLl^\dag & \iff & a \lt b \mbox{ in } \LLl\  \vee\ \bv(a)\cap \fv(b) \neq \emptyset
\eear 
where $\bv (a)$ and $\fv(b)$ are the sets of the bound and the free variables respectively, as described in section \ref{Simple typing}. The claim is that this defines a functor $\Tr:\Prg^\crl\to \Prg$, as described in the Appendix. The grading on $\Prg$ is trivial, i.e. $\lvert U\rvert = 1$ for all $U$. The normality requirement 
\begin{multline*}
\Tr^U\left(A\otimes U\tto{f} B\otimes V \tto{B\otimes u} B\otimes U\right)\ = \\ \Tr^V\left(A\otimes V\tto{A\otimes u} A\otimes U \tto{f} B\otimes V\right)
\end{multline*}
is satisfied because for
\bear
f & =&  \binp{cc}\vec{x}_A & \vec{y}_U\einp [P] \boup{cc}\vec{s}_B & \vec{t}_V \eoup \mbox{ and}\\
u & =&  \binp{c}\vec{z}_V\einp [Q] \boup{c}\vec{t}_U \eoup
\eear
it is the property of iterative algebras that the same solutions of the system
\bear
\vec{y}_U & = & \vec{r}_U(\vec{z}_V) \\
\vec{z}_V & = & \vec{t}_V(\vec{x}_A,\vec{y}_U)
\eear
are obtained both from 
\bear
\vec{y}_U & = &  \vec{r}_U\left(\vec{t}_V(\vec{x}_A,\vec{y}_U)\right)\mbox{ and from}\\ 
\vec{z}_V & = &  \vec{t}_V\left(\vec{x}_A,\vec{r}_U(\vec{z}_V)\right)
\eear
The requirement that 
\bear
\Tr^U\left(A\otimes U\tto{f\otimes U} B\otimes U\right) & = & \left(A\tto{f} B\right)
\eear
follows directly from the definition, because for $f = \binp{c}\vec{x}_A \einp [P] \boup{c}\vec{s}_B \eoup$, we have $f\otimes U = \binp{cc}\vec{x}_A & \vec{y}_U\einp [P] \boup{cc}\vec{s}_B & \vec{y}_U \eoup$ and thus $P^\dag (\vec{y}_U/\vec{y}_U) = P$ because $\vec{y}_U$ does not occur in $P$.  The final requirement, that regular scalars $\big(\big)[P]\big<\big>$ are invertible, follows from the fact that the variables in a closed cord $P$ must be bound by the $(\nu m)$ operator. The scalar denominator in a morphism $\frac{f}{s} = \frac{\left(x_A \right)[P] \left<s_B \right>}{()[P']<>}$ displays the random nonces of the cord $P$, and the equivalence $\frac{f}{s} \sim \frac{g}{t}$ identifies the processes modulo their fresh nonces. This completes the proof that the iterative structure on $\TTT$ induces a trace structure on $\Prg = \Prg_{\TTT,\WWW,\AAA}$.

The converse, that the trace operation in $\Prg$ induces an iteration
operation in $\TTT$ is proven by retracing a suitable special case of
the above construction backwards. Given a guarded system
$\vec{y}_{1..k} = \vec{s}_{1..k}(\vec{y}_{1..k})$, consider the
program $s:0+k\to k+k$ in the form
\[ \binp{c} \vec{y}_{1..k}\einp [] \boup{cc} \vec{y}_{1..k} &
\vec{s}_{1..k}(\vec{y}_{1..k}) \eoup 
\]
and use the properties of the trace to show that the program
$\Tr^k_{0k}(s)$, which is in the form $()[]< \vec{s}^{\
\dag}_{1..k}>$, gives the unique solution of the given system. 
\epr

\begin{remarkno} Note that the trace operation generally makes some
of the previously bound variables. In particular, the variables
$\vec{y}^{\ *}_{k..\ell}$, formed from the repetitions of a subset of
$\vec{y}_{k..\ell}$ that the non-guarded part of the system $\vec{y} =
\vec{s}(\vec{x},\vec{y},\vec{u},\vec{v})$ induces, are free in
$\Tr(p)$, although the tuple  $\vec{y}_{k..\ell}$ was of course bound
in $p$. This means that the polynomial subcategories $\Prg[\vec{v}]$
of $\Prg$ are generally not closed under the described trace
operation, and just support a {\em partial\/} trace operation. The
semantical significance of this will become clear later.  
\end{remarkno}


\subsection{Interactions}\label{Interactions}
An {\em interaction} is a cord process whose input and 
output interface is split into an {\em Initiator's\/} part
(i.e. the domain), and a {\em Responder's\/} part (the
codomain).\footnote{The Initiator can be viewed as the System, whereas the Responder as the Environment.} 

The category $\Intr$ of teams is obtained by applying the
$\Intt$-construction \cite{JSV} to the category $\Prg$ of
programs. The objects of $\Intr$ are pairs of arities. Since they will
usually correspond to agents, we give them names $A, B, C$ etc., and
write $A = <A_{+},A_{-}>$ and $B = <B_{+},B_{-}>$, where $A_{+},
A_{-}, B_{+}, B_{-}$ are some arities. An interaction of $A =
<A_{+},A_{-}>$ and $B = <B_{+},B_{-}>$ is described by a morphism
$p:A\to B$, which is a program $p:A_{+} \otimes B_{-} \to A_{-}\otimes
B_{+}$, usually written in the form 
{\footnotesize
\bear
\binp{c}\vec{x}_{A_+}\\
\vec{y}_{B_-}
\einp\ 
\Big[\ P\ \Big]\
\boup{c}\vec{s}_{A_+}\\
\vec{t}_{B_-}
\eoup
& = & 
\binp{cccc}x_1 & x_2 & \ldots & x_{A_+}\\
y_1 & y_2 & \ldots & y_{B_-}
\einp\ 
\Big[\ P\ \Big]\
\boup{cccc}s_1 & s_2 & \ldots & s_{A_-}\\
t_1 & t_2 & \ldots & t_{B_+}
\eoup
\eear
}
or graphically depicted as
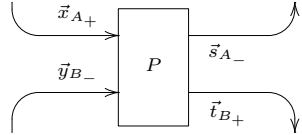
\[\def\objectstyle{\scriptstyle}
\def\g#1{\save
[].[ddddrr]!C="g#1"*[F]\frm{}\restore}
\xymatrix@C-1.9pc@R-1.8pc
{
 \ar `d[ddrrrr] [ddrrrrrr]^-{\vec{x}_{A_+}} &  &  &  &  & && &  &  & && &&  &  \\
 &&&& &  & \g1  & & & \\
 &&& & & & & &   \ar `r[rrrrrruu]_-{\vec{s}_{A_-}} [rrrrrruu]   \\
 &&& & & & & P  & & & &&&&  \\
 &&&  &  & &  &  &  \ar `r[rrrrrrdd]_-{\vec{t}_{B_+}} [rrrrrrdd] \\
 &&&  &  & & & & &  \\
 \ar `u[uurrrr] [uurrrrrr]^-{\vec{y}_{B_-}}&&&  &  & & & & & &&&&& \\
&& &  &   & & & &  & &&&&& &
}
\]
Note that the sign (polarity) of arities changes between the domain and the codomain: at the domain $A$, the input arity is $A_+$, and the output is $A_-$, while the codomain $B$ has the polarities switched, and $B_+$ is the output arity, while $B_-$ is the input. The point of this is that in the composite $p\circ q :A\to C$ of the interactions $p:A\to B$ and $q:B\to C$, represented by the programs $p:A_+ \otimes B_- \to A_-\otimes B_+$ and $q: B_+ \otimes C_- \to B_- \otimes C_+$ feeds 
\begin{itemize}
\item the $B_+$-outputs $\vec{t}_B$ of $p$ to the $B_+$ inputs $\vec{x}_B$ of $q$, and 
\item the $B_-$ outputs $\vec{s}_B$ of $q$ back to the $B_-$ inputs $\vec{y}_B$ of $p$.
\end{itemize}
Graphically, the composite $p\circ q :A\to C$ is thus
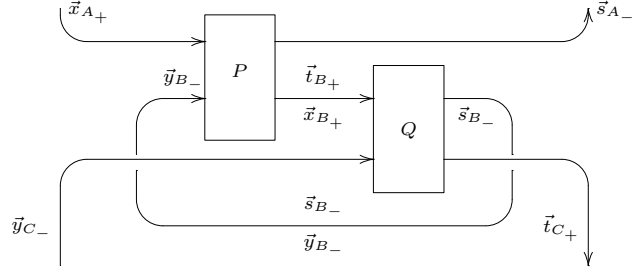
\[\def\objectstyle{\scriptstyle}
\def\g#1{\save
[].[ddddrr]!C="g#1"*[F]\frm{}\restore}
\xymatrix@C-1.9pc@R-1.8pc
{
 \ar `d[ddrrrr]^-{\vec{x}_{A_+}} [ddrrrrrrrr] &&&  &&& &&&  &&& &&& &&&  &&& &&&  &&& &&& \\
 &&&&& &&& \g1  & & & \\
 &&&&& & & & & &   \ar `r[rrrrrrrrrrrrrrrruu] [rrrrrrrrrrrrrrrruu]_-{\vec{s}_{A_-}}   \\
 &&&&& &&&   & P  && &&& && \g2 &&&  \\
 &&&&&  &&&  &&  \ar[rrrrrr]^-{\vec{t}_{B_+}}_{\vec{x}_{B_+}} &&&  &&& 	&&  \ar@{-} `r[rrrrdd]_>>{\vec{s}_{B_-}} [rrrrdd] & &&& &&& \\
&& &&&  &&& &&&  &&& &&&  Q  & & & &&&& \\
&& && \ar `u[uurr] [uurrrr]^-{\vec{y}_{B_-}} &  &&& &&&  &&& &&& & 
 \ar `r[rrrrrrrrdd] [rrrrrrrrddddd]_{\vec{t}_{C_+}} &&&& \ar@{-} `d[dddll] `[llllllllllllllllll]_-{\vec{s}_{B_-}}^-{\vec{y}_{B_-}} [llllllllllllllllll]
 \\
&&&  &&& &&&  &&&  &&&  &&& &&& &&& &&& &&&  \\
&&&  &&& &&&  &&&  &&&  &&& &&& &&& &&& &&& \\
&&&  &&& &&&  &&&  &&&  &&& &&& &&& &&& &&&  \\
&&&  &&& &&&  &&&  &&&  &&& &&& &&& &&& &&& \\
\ar `u[rrrrrrrruuuuu]^{\vec{y}_{C_-}}  [rrrrrrrrrrrrrrrruuuuu]&&&  &&& &&&  &&&  &&&  &&& &&& &&& &&& &&&
}
\]
The loop is executed using the trace operation defined in the preceding section. The syntactic view of this operation is:
{\footnotesize
\bear
\binp{ccc} x^A_1 & \ldots & x^A_{A_+}\\
y^B_1 & \ldots & y^B_{B_-} 
\einp 
& \Big[ P \Big] &
 \boup{ccc} s^A_1 & \ldots & s^A_{A_-}\\
t^B_1 & \ldots & t^B_{B_+} 
\eoup \\
& \circ & \\
\binp{ccc} x^B_1 & \ldots & x^B_{B_+}\\
y^C_1 & \ldots & y^C_{C_-} 
\einp 
& \Big[ Q \Big] &
 \boup{ccc} s^B_1 & \ldots & s^B_{B_-}\\
t^C_1 & \ldots & t^C_{C_+} 
\eoup 
\\
& = & \\
\binp{ccc} x^A_1 & \ldots & x^A_{A_+}\\
y^C_1 & \ldots & y^C_{C_-} 
\einp 
& \left[ \begin{array}{c}(P(\vec{s}^{\ \dag}_{B_-}/\vec{y}_{B_-})\otimes C_- )\ \oslash \\ (A_-\otimes Q(\vec{t}^{\ \dag}_{B_+}/\vec{x}_{B_+})\end{array} \right] &
 \boup{ccc} s^A_1 & \ldots & s^A_{A_-}(\vec{s}^{\ \dag}_{B_-}/\vec{y}_{B_-})\\
t^C_1 & \ldots & t^C_{C_+}(\vec{t}^{\ \dag}_{B_+}/\vec{x}_{B_+}) 
\eoup 
\eear
}
where $\vec{s}^{\ \dag}_{B_-}$ and $\vec{t}^{\ \dag}_{B_+}$ constitute the solution of the system:
\bear
\vec{y}_{B_-} & = & \vec{s}_{B_-}\left(\vec{x}_{B_+},\vec{y}_{C_-}\right)\\ 
\vec{x}_{B_+} & = & \vec{t}_{B_+}\left(\vec{x}_{A_+},\vec{y}_{B_-}\right)
\eear 
The fact that, by the laws of the iterative algebras, these solutions can be extracted in any order, either from
\bear
\vec{y}_{B_-} & = & \vec{s}_{B_-}\left(\vec{t}_{B_+}\left(\vec{x}_{A_+},\vec{y}_{B_-}\right),\vec{y}_{C_-}\right)\mbox{ or from}\\ 
\vec{x}_{B_+} & = & \vec{t}_{B_+}\left(\vec{x}_{A_+}, \vec{s}_{B_-}\left(\vec{x}_{B_+},\vec{y}_{C_-}\right)\right)
\eear 
implies that 
\bear
\binp{ccc} x^A_1 & \ldots & x^A_{A_+}\\
y^C_1 & \ldots & y^C_{C_-} 
\einp 
& \left[ \begin{array}{c}(P(\vec{s}^{\ \dag}_{B_-}/\vec{y}_{B_-})\otimes C_- )\ \oslash \\ (A_-\otimes Q(\vec{t}^{\ \dag}_{B_+}/\vec{x}_{B_+})\end{array} \right] &
 \boup{ccc} s^A_1 & \ldots & s^A_{A_-}(\vec{s}^{\ \dag}_{B_-}/\vec{y}_{B_-})\\
t^C_1 & \ldots & t^C_{C_+}(\vec{t}^{\ \dag}_{B_+}/\vec{x}_{B_+}) 
\eoup 
\\
& = & \\
\binp{ccc} x^A_1 & \ldots & x^A_{A_+}\\
y^C_1 & \ldots & y^C_{C_-} 
\einp 
& \left[ \begin{array}{c}(A_+\otimes Q(\vec{t}^{\ \dag}_{B_+}/\vec{x}_{B_+})\ \oslash \\ (P(\vec{s}^{\ \dag}_{B_-}/\vec{y}_{B_-})\otimes C_+ )\end{array} \right] &
 \boup{ccc} s^A_1 & \ldots & s^A_{A_-}(\vec{s}^{\ \dag}_{B_-}/\vec{y}_{B_-})\\
t^C_1 & \ldots & t^C_{C_+}(\vec{t}^{\ \dag}_{B_+}/\vec{x}_{B_+}) 
\eoup 
\eear
which corresponds to the transformation of the above graphic representation, where the $Q$-box would be moved to the left of the $P$-box along the loop. 

Note that these equivalent forms of the definition are just the unfoldings of the general formulas\footnote{For simplicity, we omit the evident commutation isomorphisms.}
\bear
p\circ q & = & \Tr^{B_-}\Big(A_+\otimes B_- \otimes C_- \tto{p \otimes C_-} A_-\otimes B_+ \otimes C_- \tto{A_- \otimes q} A_-\otimes B_-\otimes C_+\Big)\\
& = & \Tr^{B_-\otimes B_+}\Big(A_+\otimes B_- \otimes B_+ \otimes C_- \tto{p \otimes q} A_-\otimes B_+ \otimes B_-\otimes C_+ \Big) \\
& = & \Tr^{B_+}\Big(A_+\otimes B_+ \otimes C_- \tto{A_+ \otimes q} A_+\otimes B_- \otimes C_+ \tto{p \otimes C_+} A_-\otimes B_+\otimes C_-\Big)
\eear
which equivalently define the composition of $p:A_+\otimes B_-\to A_-\otimes B_+$ and $q:B_+\otimes C_-\to B_-\otimes C_+$ in the free compact category $\Intt(\CCc)$ over an arbitrary traced monoidal category $\CCc$. The
compact structure of the category $\Intr$ of interactions is thus
defined as usually in $\Intt(\CCc)$, because $\Intr =
\Intt(\Prg)$. The unit $\eta : 0 \to A^\ast \otimes A$ and the counit
$\varepsilon : A\otimes A^\ast \to 0$ both correspond to the buffer on
$A$.

The monoid of scalars consists of all cord spaces $\LLL$, taken with
both interfaces empty. The scalar multiplication is the parallel
composition.\footnote{Note that $s\circ t = \Tr^0(s\otimes t)$ gives
$s\circ t = s\otimes t$ for the scalars in $\Intr$, whereas in $\Prg$
the parallel and the sequential composition of the programs $s,t:0\to
0$ are quite different.} The embeddings 
\begin{alignat*}{5}
{\sf Init}\ :\ \Prg & \to  \Intr &\qquad\qquad\qquad && {\sf Resp}\ :\ \Prg^{op} & \to \Intr\\
n &\mapsto  <n,0>  &&& n &\mapsto  <0,n> 
\end{alignat*}
map programs respectively to the Initiator-only interactions and the
Responder-only interactions. They both display $\Intr$ as the free
compact category over $\Prg$.

If semantics for actions in $\AAA$ is given in such a way that
processes, presented as cord spaces, are reversible, with each send consumed by a single receive, and with an involution $\dag : \LLL \to \LLL$ inverting the order of actions, then the category $\Prg$ comes with an
involutive functor $\dag : \Prg^{op}\to \Prg$. This functor lifts along the $\Intt$-construction, and $\Intr$ becomes a $\dag$-compact category too, suitable for presenting, and perhaps analyzing quantum protocols \cite{Abramsky-Coecke,PavlovicD:Qabs}.

\begin{remarkno} The category $\Prg$ of cord processes has all
projections $\binp{cc}x&y\einp\ []\ \boup{c} x\eoup$ and diagonals
$\binp{c}x\einp\ []\ \boup{cc} x & x \eoup$ (albeit not natural
transformations, so $\Prg$ is not cartesian). Neither of these
families is preserved under the $\Intr$-construction, and $\Intr$ only
has the diagonals and projections for the embedded copies of the
Initiator-only and Responder-only programs.  
\end{remarkno}

%

\section{Protocols and attacks as interactions}\label{Protocols-sec}
A {\bf protocol\/} consists of a process and a
nonempty set of the desired runs. We present protocol processes as cord processes, and suggest the desired runs typographically: the local time of each cord spaces flows top down, whereas the desired interactions are aligned horizontally. --- Note that the local time in cords written as process expressions, like we did so far, flows left to right. We change this convention in this final section, in order to be able to fit a protocol on a page. 

When they participate a protocol, the agents\footnote{Recall from Sec.~\ref{Cord spaces} that the terms \emph{agent} and \emph{location} are used in cord calculus interchangeably, denoting the elements of the set $\WWW$.} play various {\em roles} in it. Formally, roles can thus be viewed as agent variables, that get instantiated to agent names when a particular agent assumes a role in a particular protocol run
\cite{PavlovicD:ESORICS04,PavlovicD:CSFW05}. 

\subsection{The Needham-Schroeder Public Key protocol (NSPK)}
\subsubsection{Prerequisites} 
To dam the flood of parentheses, we write functions in a curried
form: a function of two arguments is written $Ekx$ instead of
$E(k,x)$. This leaves $(-,-):\TTT\times \TTT \to \TTT$ to denote the
pairing operation. Formally, we assume that the agent identifiers are
terms, i.e. $\WWW\subset \TTT$, so that any operation on $\TTT$ can
also be applied on $\WWW$.

An abstract form of the Public Key Infrastructure is expressed by the
assumption that all agents given in advance the maps $E,D:\TTT\to\TTT$ and
$k:\WWW\to\TTT$, which satisfy the equation
\bea\label{decr}
D\overline{k}_X (E k_X y) & = & y
\eea
for all $X:\WWW$ and $y:\TTT$, and for map $\overline{k}:\WWW\to
\TTT$, which is not publicly known. Formally, these three maps are given as
the common context to all processes, represented as cords. In other
words, we begin from the cord category $\Prgg[E,D,k]$.

\subsubsection{The protocol}
The cord process representing the \NSPK\ protocol is:
{\footnotesize
\[
\binp{ccc} X & \overline{k}_X & Y \\
Y' &  \overline{k}_{Y'}  
\einp\
\left[\begin{array}{cc}
\new{m}_X \\
\send{E k_Y(X,m)}_X &\recv{u'}_{Y'}\\
&  \match{X',m'}{D\overline{k}_{Y'} u'}_{Y'}\\
& \new{n'}_{Y'} \\
\recv{x}_X & \send{E k_{X'}(m', n')}_{Y'}\\
\match{m,n}{D \overline{k}_X x}_X \\
\send{E k_Y n}_X & \recv{w'}_{Y'} \\
& \match{n'}{D\overline{k}_{Y'} w'}_{Y'}
\end{array}
\right]\
\boup{cccc} X & Y & m& n\\
X' & Y' & m' & n' 
\eoup
\]
}
Note again that the ordering of the cord space is now top-down,
rather than left-right; and that the spaces between the actions are
introduced to align horizontally the actions that should correspond to each other in the desired run of the protocol. When confusion seems unlikely, we write
$action_{Agent}$ instead of $a$ whenever $a$ satisfies $a \isA action_{Agent}$. For simplicity, we omit the source and destination fields from the
send and receive actions, and write e.g. $\send{t}$ instead
$\send{A\rar B:t}$.  

Semantics of actions is described in \cite{PavlovicD:CSFW05,PavlovicD:ESORICS06}. The send and the receive
actions are largely self-explanatory, as is the fresh value generation $\new{m}$. Executing a match action $\match{s_1,\ldots,s_k}{t_1,\ldots t_\ell}$,  succeeds if $k = \ell$ and for every $i$ such that $1\leq
i\leq k$,  
\begin{itemize}
\item either $s_i$ is a closed term and $s_i = t_i$,
\item or $s_i$ is a variable, and the effect of matching is the
assignment $s_i:= t_i$. 
\end{itemize}

An obvious security requirement from the \NSPK\ protocol is that for every run of
the process represented by the above cord process holds that
\begin{itemize}
\item $X=X'$ and $Y=Y'$, i.e. $X$ and $Y$ know that they share the
session, and
\item $m=m'$ and $n=n'$, i.e. they share the same values.
\end{itemize} 
For the run connecting the sends and the receives
written on the same line, this follows from the assumptions that for every
$X$ and $x$
\begin{itemize}
\item only $X$ knows $\overline{k}_X$, and3
\item the only way to extract $x$ from $Ek_X x$ is via (\ref{decr}).
\end{itemize}
A stronger security requirement is that upon the completion of a run, the
freshly created values $m$ and $n$ are {\em only\/} known to $X$
and $Y$.

\subsubsection{The attack}
However, besides the desired run, suggested above, the \NSPK\ protocol has other runs. E.g., consider the cord processes $\NSPK_1$ and $\NSPK_2$ in Fig.~\ref{figure}.
\begin{sidewaysfigure}
\bear
\NSPK_1\hspace{1em} = \hspace{6em} 
\binp{ccc} X & \overline{k}_X & Z \\
Z' &  \overline{k}_{Z'} & z' 
\einp &
\left[\begin{array}{cc}
\new{m}_X \\
\send{E k_Z(X,m)}_X &\recv{u'}_{Z'}\\
&  \match{X',m'}{D\overline{k}_{Z'} u'}_{Z'}\\
\recv{x}_X & \send{E k_{X'}z'}_{Z'}\\
\match{m,n}{D \overline{k}_X x}_X \\
\send{E k_Z n}_X & \recv{w'}_{Z'} \\
& \match{n'}{D\overline{k}_{Z'} w'}_{Z'}
\end{array}
\right] &
\boup{cccccc} X & Z & m& n\\
X' & Z' & m' & n' & \overline{k}_{Z'}  & Y' 
\eoup
\\
\\
\\
\\
\NSPK_2\ = \ \ \ 
\binp{cccccc} X' & Z' & m' & n' & \overline{k}_{Z'}  & Y'  \\
Y'' &  \overline{k}_{Y''}  
\einp &
\left[\begin{array}{cc}
\send{E k_{Y'}(X',m')}_{Z'} &\recv{u''}_{Y''}\\
&  \match{X'',m''}{D\overline{k}_{Y''} u''}_{Y''}\\
& \new{n''}_{Y''}\\ 
\recv{z'}_{Z'} & \send{E k_{X''}(m'', n'')}_{Y''}\\
\send{E k_{Y'} n'}_Z & \recv{w''}_{Y''} \\
& \match{n''}{D\overline{k}_{Y''} w''}_{Y''}
\end{array}
\right] &
\boup{ccccc} Z' &  \overline{k}_{Z'} & z' \\
X" & Y'' & m'' & n'' 
\eoup
\eear
\vspace{1.2\baselineskip}
\bear
\mathbf{\NSPK_1\circ \NSPK_2}\hspace{2em} = \hspace{3em}
\binp{ccc} X & \overline{k}_X & Z \\
Y'' &  \overline{k}_{Y''} &  
\einp &
\left[\begin{array}{ccc}
\new{m}_X \\
\send{E k_Z(X,m)}_X &\recv{u'}_{Z'}\\
&  \match{X',m'}{D\overline{k}_{Z'} u'}_{Z'}\\
&\send{E k_{Y'}(X',m')}_{Z'} &\recv{u''}_{Y''}\\
& &  \match{X'',m''}{D\overline{k}_{Y''} u''}_{Y''}\\
& & \new{n''}_{Z''} \\
& \recv{z'}_{Z'} & \send{E k_{X''}(m'', n'')}_{Y''}\\
\recv{x}_X & \send{E k_{X'} z'}_{Z'}\\
\match{m,n}{D \overline{k}_X x}_X \\
\send{E k_Z n}_X & \recv{w'}_{Z'} \\
& \match{n'}{D\overline{k}_{Z'} w'}_{Z'} \\
& \send{E k_{Y'} n'}_{Z'} & \recv{w''}_{Y''} \\
&& \match{n''}{D\overline{k}_{Y''} w''}_{Y''}
\end{array}
\right] &
\boup{cccc} X & Z & m& n\\
X'' & Y'' & m'' & n'' 
\eoup
\eear
\caption{The attack components and their composition}
\label{figure}
\end{sidewaysfigure}
They are derived from the \NSPK\ by modifying in $\NSPK_1$ the
Responder, and in $\NSPK_2$ the Initiator. In both cases, instead of
generating a fresh value, the agent takes it from the input
interface. Moreover, the challenge received from the peer is forwarded
to the output interface. To compose $\NSPK_1$ and $\NSPK_2$, we proceed as in section \ref{Interactions}
\begin{itemize}
\item connect the Responder interfaces of $\NSPK_1$ to the Initiator interfaces of $\NSPK_2$, as suggested by the chosen names
\item extend the parallel composition of $\NSPK_1$ and $\NSPK_2$ by the ordering imposed by condition (\ref{lt}), as follows:
\begin{itemize}
\item $X',m' \in  \bv\match{X',m'}{D\overline{k}_{Z'}u'}\cap \fv\send{Ek_{Y'}(X',m')}\\ \Longrightarrow \match{X',m'}{D\overline{k}_{Z'}u'}^1_{Z'}\lt \send{Ek_{Y'}(X',m')}^2_{Z'} $
\item $z' \in \bv\recv{z'}\cap \fv\send{Ek_{X'}z'}  \\ \Longrightarrow  \recv{z'}^2_{Z'}\lt \send{Ek_{X'}z'}^1_{Z'} $,
\item $n' \in \bv\match{n'}{D\overline{k}_{Z'}w'}\cap  \fv\send{Ek_{Y'}n'} \\ \Longrightarrow  \match{n'}{D\overline{k}_{Z'}w'}^1_{Z'}\lt \send{Ek_{Y'}n'}^2_{Z'} $
\end{itemize}
\end{itemize}
where the superscript $(-)^1$ denotes the actions of $\NSPK_1$ and $(-)^2$ the actions of $\NSPK_2$. The resulting interaction $\NSPK_1\circ \NSPK_2$ is displayed in Fig.~\ref{figure}.
Upon the termination of the run of the resulting cord process, $m=m''$
and $n=n''$ will hold, as well as $X=X''$, but $Z\neq Y''$. This means that the requirement that $X$ and $Y$ share the session with each
other is not satisfied, since $X$ thinks that she shares it with
$Z\neq Y$, whereas $Y$ thinks she shares it with $X=X''$. Moreover,
$Z$ knows both freshly generated values $m$ and $n$, and they are thus not secret between $X$ and $Y$.

\section{Discussion and future work}\label{Discussion-sec}
We provided a general categorical view of the MM attacks, and instantiated it on the NSPK protocol. Although the trace structure and the coalgebraic nature of the MM interactions has been displayed, this categorical view did not turn out to be as simple or as succinct as one would like. This may be due to the cord calculus infrastructure, which was originally designed for use in a software tool \cite{PavlovicD:ARSPA06}, and later adapted for human consumption. While a certain amount of verbosity may be unavoidable in security formalisms, there is hope that the incremental approach will lead to more convenient languages \cite{PavlovicD:ICDCIT12}. The shortcomings of the underlying process calculus notwithstanding, the presented categorical constructions seem to substantiate the idea that \emph{hiding}, inherent in MM, can be captured using the monoidal trace structure. The most important technical features of the presented categorical analysis seem to be that
\begin{itemize}
\item the data flows resulting from the interactions correspond to the \emph{iterative structure}, which resolves the systems of equations induced by the interactions, and thus effectively propagates the terms sent in messages;
\item the functorial view of the \emph{normal} traces over the cord category $\Tr : \Prg^\circlearrowleft \to \Prg$, arising from this iterative structure, can be used to analyze the possible MM attacks, since all hidden interactions that result in a process $f$ observable in $\Prg$ can be found in its inverse image $\Tr^{-1}(f)$ in $\Prg^\circlearrowleft$.
\end{itemize}
As intriguing as they may be, both these features are clearly beyond the scope of the present paper (even with its swollen Appendices). If the approach turns out to be effective, the intended next step, as mentioned in the Introduction, would be to try to formalize \emph{chosen protocol attacks} \cite{Kelsey-Schneier-Wagner}. To add more intrigue to the story, this goal seems to require two monoidal structures, to allow distinguishing the situation
\begin{itemize}
\item $A\otimes B$, where the roles $A$ and $B$ are played by the same principal, who controls and can mix all information sent and received in both roles; from the situation
\item $A\oplus B$, where the roles $A$ and $B$ only exchange information through messages.
\end{itemize}
Interestingly, the trace structure, at least in its normal flavor, does seem to have a natural generalization in such a framework, as well as a corresponding $\Intt$-construction, capturing the MM-interactions that naturally evolve there.

\subsubsection{Acknowledgement.} This work was started some 10 years ago, as a collaboration with Samson Abramsky, who had introduced me to semantics of interaction a bit earlier, and to computer science just before that. His influence on the presented ideas cannot be overestimated. On the other hand, as our interactions extended not only across the broad areas of common interest, but also across the great distances that separated us, all the shortcomings of the presented text remain entirely my responsibility. Cord calculus was developed in joint work Matthias Anlauff, Iliano Cervesato, Anupam Datta, Ante Derek, Nancy Durgin, Cathy Meadows, and John Mitchell.

\bibliography{CIM-ref,PavlovicD}
\bibliographystyle{plain}

\appendix
\section{Appendix: Traces over graded categories}

For simplicity, and without loss of generality, we assume that the monoidal structures that we consider are strictly associative and unitary, i.e. $(A\otimes B)\otimes C = A\otimes (B\otimes C)$ and $A\otimes I = I\otimes A = A$. 

\subsection{Graded categories and loop categories}
\begin{definition}A small symmetric monoidal category 
\[ \CCc\times \CCc \stackrel{\otimes}{\to}\CCc
\stackrel{I}{\longleftarrow} 1\]
is said to be \emph{graded} by a monoid homomorphism 
\begin{eqnarray*}
 (\CCc,\otimes,I) & \stackrel{|-|}\to & (\IIi,\circ,1)
\end{eqnarray*}
where $\IIi = \CCc(I,I)$, $1 = {\rm id}_I$. The elements of 
\bear
\IIi^*  & = &  \{s\in \IIi\ |\ \forall t\in \IIi\exists u\in \IIi.\ stu \neq st\}
\eear are called {\em regular}. A graded symmetric monoidal category is called {\em local monoidal\/} if all of its regular scalars are invertible.
\end{definition}

\begin{definition}
For a local monoidal category $\CCc$, the trace structure in the sense of \cite{JSV} is said to be \emph{normal} if in addition to the standard axioms the trace operation also satisfies the \emph{normality} requirement:
\bear
\Tr^U\left(A\otimes U\tto{f\otimes U} B\otimes U\right) & = & \left(A\tto{f} B\right)
\eear
\end{definition}

\subsubsection{Remark.}
The trace structures with respect to both the additive and the multiplicative monoidal structure of the category relations are normal. On the other hand, the standard trace structure over the category of vector spaces is not normal. The upshot of the normality requirement is that it opens a functorial view of the traces.

\subsubsection{Loop category.}
Given a graded category $\CCc$, let $\CCc^\circlearrowleft$ be the category defined
\begin{eqnarray*}
|\CCc^\circlearrowleft | & = & |\CCc|  \\
\CCc^\circlearrowleft (A,B) & = & \Big(\sum_{U\in |\CCc|}\ \CCc(A\otimes U, B\otimes U)\ \times\ \IIi^*\Big)\ \Big/ \boldsymbol{\sim} 
\end{eqnarray*}
A $\CCc^\circlearrowleft$-morphism from $A$ to $B$ is thus an equivalence class of pairs $<f,s>$, where $f:A\otimes U\to B\otimes U$ is a $\CCc$-morphism and $s:I\to I$ is a regular scalar. Writing such pairs as fractions $\frac f s$, we define $\sim$ as the smallest equivalence relation containing the following relations
\begin{itemize}
\item the coend equivalence \cite[IX.6] {MacLane:CWM}
\bear
\xymatrix@R-.3pc@C-1pc
{ & \CCc(A\otimes U, B\otimes U) \ar@{-->}[rrrrd] \\
\CCc(A\otimes V, B\otimes U)
\ar@{-->}[rrrrr]
\ar[ur]^{(-)\circ (A\otimes u)} 
\ar[dr]_{(B\otimes u)\circ (-)} 
&&&&& \CCc^\circlearrowleft (A,B)\\
& \CCc(A\otimes V, B\otimes V) \ar@{-->}[rrrru]
}
\eear
which means 
\bear
\frac{(B\otimes u)\circ f}{s} &   \boldsymbol{\sim} & \frac{f\circ(A\otimes u)}{s}
\eear
\[
\xymatrix@R-.5pc@C-.5pc{
& A\otimes U \ar[dl]_f && A\otimes V \ar[dr]^{A\otimes u} \\
B\otimes V \ar[dr]_{B\otimes u} && 
&& A \otimes U
\ar[dl]^f \\
& B\otimes U && B\otimes V}
\]
\item tensor is normalized
\bear
\frac{f\otimes U}{s\circ\lvert U\rvert}\hspace{1.5em} & \boldsymbol{\sim} & \hspace{1.5em}  \frac{f}{s} 
\eear
\[
\xymatrix@R-1pc@C-1pc{
A\otimes U \ar[d]_{f\otimes U} && A \ar[d]^{f} \\ 
B\otimes U && B}
\]
\item regular scalars are invertible in all morphisms
\bear
\frac{f}{s} \boldsymbol{\sim} \frac{g}{t}&\iff &
\exists uv\in \IIi^*.\ \ u\circ f = v\circ g \wedge u\circ s = v\circ t 
\eear
\end{itemize}
We proceed to define the categorical structure of $\CCc^\circlearrowleft$. Given
\begin{itemize}
\item $f\in \CCc^\circlearrowleft(A,B)$ as  $A\otimes
U\tto{f_0/f_1} B\otimes U$, 
\item $g\in \CCc^\circlearrowleft(B,C)$ as $B\otimes V\tto{g_0/g_1} C\otimes V$, and
\item $h\in \CCc^\circlearrowleft(C,D)$ as $C\otimes V\tto{h_0/h_1} D\otimes V$
\end{itemize}
then the \textbf{composition} $f\circ g\in \CCc^\circlearrowleft(A,C)$ can be viewed  as
\[
\xymatrix@C-.8pc{
A\otimes U\otimes V \ar[rr]_{\frac{f_0\otimes V}{f_1\circ |V|}} && B\otimes U\otimes V
\ar[d]|{B\otimes c} && C\otimes U\otimes V\\
&& B\otimes V\otimes U \ar[rr]_{\frac{g_0\otimes U}{g_1\circ |U|}} && C\otimes V\otimes U
\ar[u]|{C\otimes c}
}
\]
or equivalently
\[
\xymatrix@C-.8pc{
A\otimes U\otimes V \ar[rr]_{\frac{f_0\otimes V}{f_1\circ |V|}} && B\otimes U\otimes V
\ar[d]|{B\otimes c}\\
A\otimes V\otimes U \ar[u]|{A\otimes c} && B\otimes V\otimes U
\ar[rr]_{\frac{g_0\otimes U}{g_1\circ |U|}} && C\otimes V\otimes U
}
\]
whereas the \textbf{tensor} $f\otimes h\in \CCc^\circlearrowleft(A\otimes C,B\otimes D)$ is
\[
\xymatrix{
A\otimes C\otimes U \otimes V \ar[d]|{A\otimes c\otimes V} && B\otimes D\otimes U\otimes V \\ 
A\otimes U\otimes C \otimes V \ar[rr]_{\frac{f_0\otimes h_0}{f_1\circ h_1}} && B\otimes U\otimes D \otimes  V
\ar[u]|{B\otimes c\otimes V}
}
\]
Since the scalars in $\CCc^\circlearrowleft$ are the fractions of those in $\CCc$, the \textbf{grading} of $\CCc^\circlearrowleft$ is inherited from $\CCc$. Finally, the \textbf{trace} operation is
\[
\prooftree
\hspace{2em}f \ = \ \frac{(A\otimes U)\otimes V  \tto{f_0}  (B\otimes U)\otimes V}{f_1}\ \in\  \CCc^\circlearrowleft(A\otimes U,B\otimes U)
\justifies
\Tr^U_{AB} f\ = \ \frac{A\otimes (U\otimes V)  \tto{f_0}  B\otimes (U\otimes V)}{f_1}\  \in \ \CCc^\circlearrowleft(A,B)\hspace{3.8em}
\thickness=.1em
\endprooftree
\]
To see that the operators $\Tr^U_{AB} : \CCc^\circlearrowleft(A\otimes U,B\otimes U) \to \CCc^\circlearrowleft(A,B)$ satisfy the trace axioms from \cite{JSV}, observe that
\begin{itemize}
\item dinaturality (sliding) and yanking laws are imposed by the definition of $\boldsymbol{\sim}$
\item naturality (tightening) by the definition of the composition in $\CCc^\circlearrowleft$, whereas
\item vanishing and superposition are easily checked by inspection.
\end{itemize}

\begin{theorem}
The loop category $\CCc^\circlearrowleft$ is the free normal traced category generated by the graded monoidal category $\CCc$. Normal traced categories correspond to strong algebras $T : \CCc^\circlearrowleft \to \CCc$ for the monad $\circlearrowleft : \GGG\MMM \to \GGG\MMM$ on the category $\GGG\MMM$ of small graded categories with the grade preserving monoidal functors. The monad structure is
\begin{align*}
\boldsymbol{\eta}_\CCc\  : \  \CCc & \to  \CCc^\circlearrowleft\\
{(A\stackrel{f}{\rightarrow} B)} & \longmapsto  { \left[A\otimes I\stackrel{f\otimes I}{\rightarrow} B\otimes I\right]_\sim}\\[+2ex]
\boldsymbol{\mu}_\CCc\  : \  \CCc^{\circlearrowleft\circlearrowleft} & \to   \CCc^\circlearrowleft \\
{\left[\frac{\left[\frac{(A\otimes U)\otimes V\stackrel{f_0}{\rightarrow} (B\otimes U)\otimes V}{s}\right]_\sim}{t} \right]_\sim} & \longmapsto {\left[\frac{A\otimes (U\otimes V)\stackrel{f_0}{\rightarrow} B\otimes (U\otimes V)}{s\circ t}\right]_\sim}
\end{align*}
\end{theorem}
Towards the \textbf{proof}, observe that the arrow part of a loop algebra $T : \CCc^\circlearrowleft \to \CCc$ yields a map
\beq\label{above} \sum_{U\in \CCc} \CCc(A\otimes U,B\otimes U) 
\xymatrix@C-.5pc{\ar@{->>}[r]&} \CCc^\circlearrowleft(A, B) \xymatrix{\ar[r]^-{T_{AB}} &\CCc(A,B)
}\eeq
which boils down to a family of trace operators $$\left\{\Tr_{AB}^U : \CCc(A\otimes U,B\otimes U)\to \CCc(A,B)\right\}_{U\in \CCc}$$ Note that this is not a natural family, since precomposing on the left with $g\otimes U$ corresponds on the right to $g\circ |U|$. The operators $\Tr_{AB}^U : \CCc(A\otimes U,B\otimes U)\to \CCc(A,B)$ do satisfy the trace axioms of \cite{JSV} because:
\begin{itemize}
\item naturalities, yanking, normality $\iff$\  factoring through $\CCc^\circlearrowleft(A, B)$, 
\item superposition  $\iff \ T\circ \boldsymbol{\eta}_\CCc = id_\CCc$
\item vanishing $\iff\  T\circ \boldsymbol{\mu}_\CCc = T\circ T^\circlearrowleft$
\end{itemize}

\begin{remarkno}
Although $\circlearrowleft$ is not a KZ-monad, its restriction to symmetric monoidal posets, i.e. to ordered abelian monoids, is an idempotent monad. An ordered abelian monoid has a trace if and only if the monoid operation is an order isomorphism, i.e. $\exists x. a+x = b+x \Longrightarrow a=b$. The $\Intt$-construction generates the ordered abelian groups, since $0\leq a+a^\ast$  and $a^\ast + a\leq 0$  mean that $a^\ast=-a$.

\end{remarkno}

\section{Appendix: Uniform traces over graded categories}
\begin{definition}{\cite{HasegawaM:uniformity}}
A trace operator is {\em uniform} if 
\bear
\Tr_{AB}^U(f) & = & \Tr_{AB}^V(g)
\eear
holds whenever some $h$ makes the following diagram commute.
\[\xymatrix@C-.8pc@R-1pc{
A\otimes U \ar[dd]_f  \ar[rr]^{A\otimes h} && A\otimes V \ar[dd]^g\\ \\
B\otimes U \ar[rr]_{B\otimes h} && B\otimes V 
}
\]
\end{definition}

\subsubsection{Uniform loop category.}
Given a graded category $\CCc$, let $\CCc^\circlearrowleft$ be the category defined
\begin{eqnarray*}
|\CCc^\looparrowleft | & = & |\CCc| \\
\CCc^\looparrowleft (A,B) & = & \Big(\sum_{U\in |\CCc|}\ \CCc(A\otimes U, B\otimes U)\times \IIi^\ast \Big)\ \Big/ \boldsymbol{\approx} 
\end{eqnarray*}
where $\boldsymbol{\approx}$ extends $\boldsymbol \sim$ by the following extension of the coend equivalence:
\[
\xymatrix@R-.3pc@C-1.4pc
{ & \CCc(A\otimes U, B\otimes U) \ar@{-->}[rrrrd] \ar[dr]_{(B\otimes h)\circ (-)} \\
\CCc(A\otimes V, B\otimes U)
\ar[ur]^{(-)\circ (A\otimes u)} 
\ar[dr]_{(B\otimes u)\circ (-)} 
&& \CCc(A\otimes U, B\otimes V) \ar@{-->}[rrr]
&&& \CCc^\looparrowleft (A,B)\\
& \CCc(A\otimes V, B\otimes V) \ar[ur]^{(-)\circ (A\otimes h)} 
\ar@{-->}[rrrru]
}
\]
i.e. by adding
\[
\xymatrix@R-.5pc@C-.8pc
{ A\otimes U \ar[dd]^f \ar@{-->}[rr]|{A\otimes h}&& A\otimes V \ar[dd]_{g} \\
& \boldsymbol{\approx} \\
B\otimes U \ar@{-->}[rr]|{B\otimes h} && B\otimes V
}
\]

\begin{theorem}
The uniform loop category $\CCc^\looparrowleft$ is the free normal uniformly traced category generated by the graded monoidal category $\CCc$. Normal uniformly traced categories correspond to the strong algebras $T : \CCc^\looparrowleft \to \CCc$ for the monad $\looparrowleft : \GGG\MMM \to \GGG\MMM$ on the category $\GGG\MMM$ of small graded categories with the grade preserving monoidal functors.
\end{theorem}

\section{Appendix: Traced clones}
The monadic view of normal traced categories allows effective calculations of the trace structures. For instance, consider the monoid of natural numbers
\[ \NNn\times \NNn \stackrel{+}{\longrightarrow}\NNn
\stackrel{0}{\longleftarrow} 1\] 
as the category of sets $n = \{0,1,\ldots,n-1\}$ and functions between them. The grading is trivial. Then 
\bear
\NNn^\circlearrowleft (a,b) & = & \sum_{u\in \NNn} \big\{a+u \stackrel{f}\rightarrow b+u \ |\ \forall y\in u\exists x.\ f(x)=y\ \wedge\ (x\in u\ \vee f(y)\in u)\big\} \\
\NNn^\looparrowleft (a,b) & = &  \sum_{u\in \NNn} \big\{a+u \stackrel{f}\rightarrow b+u\  \in \NNn^\circlearrowleft (a,b)\ | \  \forall y\in u.\ f(y)=y \ \vee\ \exists i.\ f^i(y)\in b\big\}
\eear
These constructions extend to the situations when $\NNn$ is extended by algebraic operations and actions.

\begin{definition}{\cite{LawvereFW:funsem}}
Given an algebraic theory $\TTT = <\Sigma_\TTT, E_\TTT>$, where $\Sigma = \Sigma_\TTT$ is a signature, and $E = E_\TTT$ is a set of equations, the induced {\em clone} $\NNn_\TTT = \NNn[\Sigma; E]$\footnote{The notation echoes \cite{PavlovicD:MSCS97}, where free constructions over monoidal categories were analyzed as polynomial extensions.} is the category 
\bear 
\left| \NNn_\TTT\right| & = & |\NNn|\\
\NNn_\TTT(m,n) & = & \big\{\ (x_1,\ldots, x_n)<\varphi_1,\ldots,\varphi_m>\  \big\}\ \big/ \alpha
\eear 
i.e. obtained by 
\begin{itemize}
\item adjoining to $\NNn$ an arrow $m\stackrel{\varphi}{\rightarrow} n$ for every $m$-tuple $\left<\varphi_i(x_1,\ldots x_n)\right>_{i\leq m}$ of well-formed $\Sigma$-operations, modulo the $\alpha$-conversion, i.e. variable renaming; and then by
\item imposing the equations of $E$ on the obtained category.
\end{itemize}
\end{definition}

\begin{definition}\label{iterative}
An algebraic theory $\TTT$ is {\em iterative} if every system
\bear
y_1 & = & f_1(y_1,y_2,\ldots,y_k,\ldots,y_\ell)\\
y_2 & = & f_2(y_1,y_2,\ldots,y_k,\ldots,y_\ell)\\
& \cdots & \\
y_k & = & f_k(y_1,y_2,\ldots,y_k,\ldots,y_\ell) 
\eear
has a unique solution 
\bear
f^\dag_{1}(y_{k+1},\ldots,y_\ell) & = & f_1(f^\dag_{1},f^\dag_{2},\ldots,f^\dag_{k},\ldots,y_\ell)\\
f^\dag_{2}(y_{k+1},\ldots,y_\ell) & = & f_2(f^\dag_{1},f^\dag_{2},\ldots,f^\dag_{k},\ldots,y_\ell)\\
& \cdots & \\
f^\dag_{k}(y_{k+1},\ldots,y_\ell) & = & f_k(f^\dag_{1},f^\dag_{2},\ldots,f^\dag_{k},\ldots,y_\ell) 
\eear
provided that all equations are {\em guarded\/}, i.e. that none of the operations $f_j$ is a projection.
\end{definition}

\begin{theorem}
The uniform traces over the clone $\NNn_\TTT$ are in one to one correspondence with the iterative structures over the algebraic theory $\TTT$.
\end{theorem}

This is where Prop.~\ref{proposition} picks up the thread, with 
\bear 
\left| \Prg_{\TTT,\WWW,\AAA}\right| & = & |\NNn|^2\\
\Prg_{\TTT,\WWW,\AAA}\left(<k,m>,<\ell,n>\right) & = & \big\{\ (\vec{X}_{1..k},\vec{x}_{1..n})[P]<\vec{A}_{1..\ell}, \vec{\boldsymbol \varphi}_{1..m}>\  \big\}\ \big/ \alpha
\eear 
where $A_1,\ldots, A_\ell \in \AAA$, $\varphi_1,\ldots,\varphi_m\in \TTT$ and $P$ are the processes built from the actions in $\AAA$ over the locations in $\WWW$.

\end{document}